# Investigation of the near-wake behaviour of a utility-scale wind turbine

**Aliza Abraham, Teja Dasari and Jiarong Hong**[1]

St. Anthony Falls Laboratory and Department of Mechanical Engineering, University of Minnesota, Minneapolis, MN, USA

[1]Author to whom correspondence should be addressed: jhong@umn.edu

**Abstract**. Super-large-scale particle image velocimetry and flow visualization with natural snowfall is used to collect and analyse multiple datasets in the near wake of a 2.5 MW wind turbine. Each dataset captures the full vertical span of the wake from a different perspective. Together, these datasets compose a three-dimensional picture of the near-wake flow, including the effect of the tower and hub and the variation of instantaneous wake expansion in response to changes in turbine operation. A region of high-speed flow is observed directly behind the hub, and a region of low-speed flow appears behind the tower. Additionally, the hub produces a region of enhanced turbulence in its wake while the tower reduces turbulence near the ground as it breaks up turbulent structures in the boundary layer. Analysis of the instantaneous wake behaviour reveals variations in wake expansion – and even periods of wake contraction – occurring in response to changes in angle of attack and blade pitch gradient. This behaviour is found to depend on the region of operation of the turbine. These findings can be incorporated into wake models and advanced control algorithms for wind farm optimization and can be used to validate wind turbine wake simulations.

## 1. Introduction

Understanding the wind turbine wake is crucial for improving the efficiency of wind farms, as wake loss can cause power losses of 10-20% in large wind farms [1] and can increase fatigue loading on downwind turbines [2]. The behavior in the near wake (1-4 rotor diameters downstream) influences far-wake development, as the breakdown of near-wake coherent structures enhances mixing and recovery [3], and the interaction between the vortices shed from the hub and blades affects meandering [4]. This near-wake behavior is significantly influenced by changes in turbine operation. For example, when the turbine is operating with yaw error (misalignment between the wind direction and rotor direction), a spanwise force is exerted on the wake, causing it to deflect in the spanwise direction [5, 6]. Additionally, a number of studies have investigated the regulation of wake velocity deficit using blade pitch and generator torque to maximize wind farm power generation [7, 8].

Though these studies have yielded many useful results, they were all conducted using either numerical simulation or conventional field measurement techniques such as LiDAR, which have substantial limitations in resolving the complex interaction between the turbulent atmospheric flow and the turbine structure in the near wake. Specifically, simulations largely rely on simplified physics and turbine geometry while the LiDAR measurements do not have sufficient spatiotemporal resolution to capture the dynamics of rich coherent structures in the wake. On the other hand, super-large-scale particle image velocimetry (SLPIV) and flow visualization, first implemented by Hong *et al* [10], provides high-resolution velocity fields for atmospheric flows using natural snowflakes as flow tracers. This technique was first validated and applied to preliminary utility-scale wind turbine wake measurements by Hong *et al* [9] and Toloui *et al* [10]. Nemes *et al* [11] and Heisel *et al* [12] then further applied it to atmospheric boundary layer studies. Dasari *et al* [13] expanded the capabilities of the technique to a field of view > 120 m to quantify the near-wake velocity deficit and coherent blade tip vortex behaviour. Particularly, the work was the first to observe the occurrence of increased wake velocity indicative of wake contraction behind a field-scale turbine associated with changes in blade pitch. Subsequently, Abraham *et al* [14] applied SLPIV to investigate the effect of the nacelle and tower on the wake, and Abraham and Hong [15] explored the dynamic effects of turbine operation. The goal of the current paper is to combine multiple fields of view to develop a full three-dimensional understanding of novel wake behaviours observed using SLPIV.

## 2. Methods

This study was conducted at the Eolos field site in Rosemount, MN, home of a 2.5 MW 3-bladed, horizontal axis, pitch-regulated wind turbine, fully instrumented with a supervisory control and data acquisition (SCADA) system and strain gages along the blades and base of the tower. The turbine rotor has a 96 m diameter and the hub height is 80 m. The site also includes a 130 m meteorological tower (referred to as the met tower hereafter) located 170 m south of the turbine for atmospheric data collection. The site and instrumentation are described in more detail in Hong *et al* [9], Toloui *et al* [10], and Dasari *et al* [13].

The current study combines data collected during three different SLPIV deployments. Each dataset has a field of view (FOV) capturing the full vertical rotor span at a different position in the near wake: parallel to the flow and offset from the central tower plane, parallel to the flow and directly behind the tower, and perpendicular to the flow and directly behind the tower. The detailed experimental setup of each deployment is shown in figure 1, and the relevant geometric parameters are summarized in table 1. The atmospheric and turbine operational conditions during each deployment are listed in table 2.

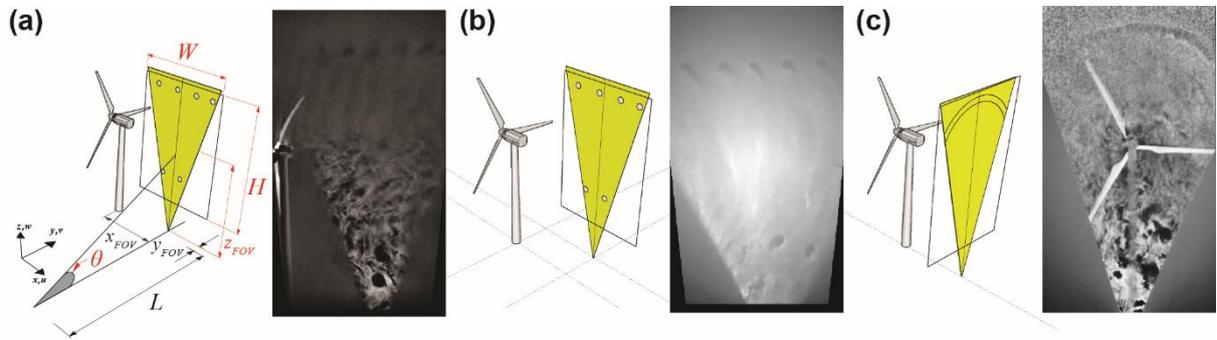

**Figure 1.** (a) Experimental setup and sample image of the dataset with a field-of-view aligned parallel to the flow directly behind the tower. Values for the parameters shown in the schematic are listed in Table 1. (b) Experimental setup and sample image of the dataset with a field-of-view aligned parallel to the flow offset from the plane of the tower. (c) Experimental setup and sample image of the dataset with a field-of-view aligned perpendicular to the flow directly behind the tower.

**Table 1.** Geometric parameters for the experimental setups of the datasets used in this study.

| Deployment date | Dataset duration | Alignment to flow | FOV location $(x_{FOV}, y_{FOV})$ | FOV elevation $(z_{FOV})$ | FOV size $(H \times W)$ | Camera-to-light distance | Tilt angle |
|---|---|---|---|---|---|---|---|
| Feb. 2, 2016 | 30 min | Parallel | $0.41D$, $0.19D$ | 81 m | 115 m × 66 m | 151 m | 27.7° |
| Mar. 12, 2017 | 62 min | Parallel | $0.35D$, $0.06D$ | 80 m | 125 m × 70 m | 171 m | 24.5° |
| Jan. 22, 2018 | 120 min | Normal | $0.19D$, $0.02D$ | 69 m | 129 m × 73 m | 166 m | 22.6° |

**Table 2.** Atmospheric and turbine operation conditions during the datasets used in this study

| Deployment date | Mean wind speed at hub height | Mean wind direction (clockwise from North) | Turbulence intensity | Mean temperature | Turbine operational region | Tip-speed-ratio |
|---|---|---|---|---|---|---|
| Feb. 2, 2016 | 10.0 m/s | 15° | 0.18 | -3.2°C | 2-3 | 5.5-7.5 |
| Mar. 12, 2017 | 5.9 m/s | 56° | 0.18 | -8.1°C | 1.5-2 | 8-11.5 |
| Jan. 22, 2018 | 9.1 m/s | 6° | 0.30 | -3.7°C | 2-3 | 5-13 |

Wake velocity fields are calculated for both datasets with fields of view aligned parallel to the flow. First, the raw images are enhanced to increase the signal-to-noise ratio and de-warped to account for the tilt angle of the camera. Then vectors are calculated using the adaptive multi-pass cross correlation algorithm from *LaVision Davis 8*. Due to the large viewing distances of the camera, the large-scale patterns formed by the snowflakes and coherent structures in the near wake are tracked to quantify the flow field, rather than individual snowflake particles, as described in detail by Dasari *et al* [13]. The resulting spatial and temporal resolution are 4 m/vector and 6 Hz respectively.

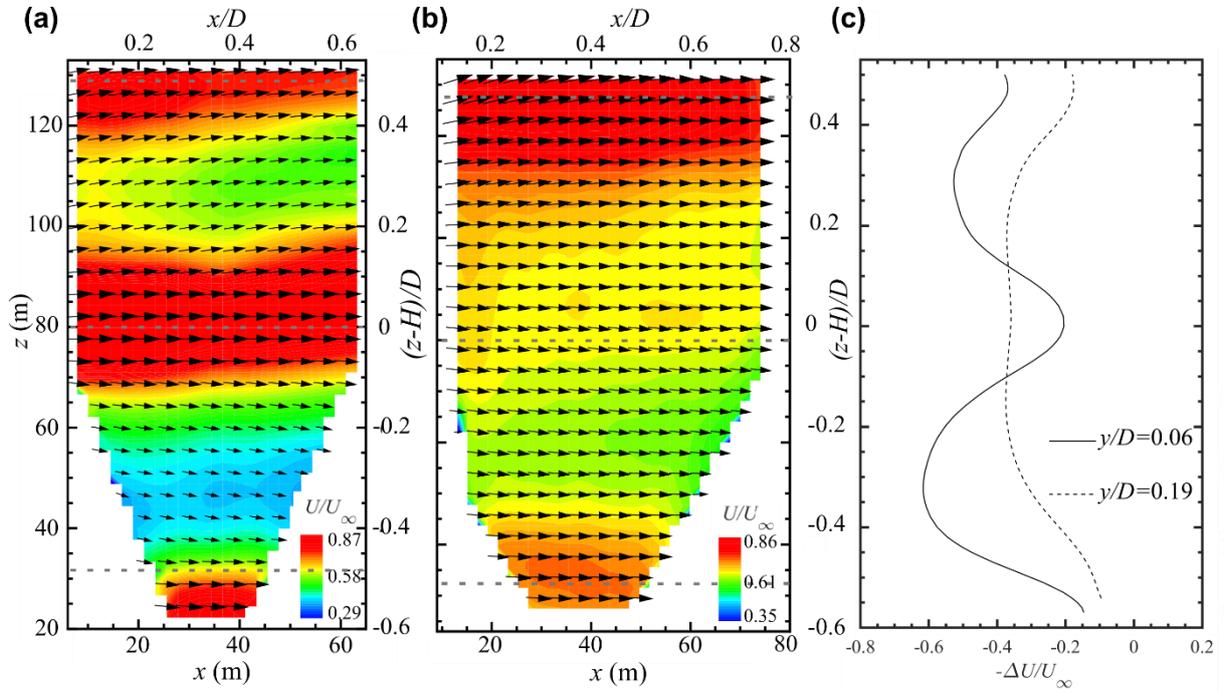

**Figure 2.** (a) Mean wake velocity field in the plane aligned with the tower. Dashed horizontal lines indicate hub and blade tip heights. (b) Mean wake velocity field in the plane offset from the tower. (c) Comparison of the velocity deficit profiles taken at $x/D = 0.41$ for measurements at the spanwise locations $y/D = 0.06$ and $y/D = 0.19$.

## 3. Results
*3.1. Mean flow*
The mean wake velocity field at the two different spanwise locations are compared in figure 2. Figure 2(a) shows the velocity field on the plane aligned with the tower. In this dataset, the wind direction fluctuates significantly, causing misalignment between the light sheet direction and wind direction. Therefore, conditional sampling is applied to select the data where the wind direction is well-aligned with the tower (misalignment angle $\Upsilon_{LW} \leq 10°$) in order to ensure the tower and hub effects are captured. In this plane, a clear velocity deficit is observed behind the blades, while the velocity deficit behind the hub is significantly reduced due to the reduction of lift at the blade root. The velocity deficit is stronger behind the tower due to the increased blockage. Figure 2(b) shows the wake velocity field on a plane offset $0.19D$ from the tower in the spanwise direction. At this location, the features of the tower and hub are no longer visible. The velocity deficit is relatively uniform throughout the span of the wake. In figure 2(c), the velocity deficit profiles taken at $x = 0.41D$ from each spanwise plane are compared. On the tower plane, the profile has a strong double-gaussian shape due to the presence of the hub, while on the off-tower plane, only 12 m away, the profile is nearly gaussian. This highlights the limited spanwise extent of the hub wake.

The in-plane turbulent kinetic energy of the wake, defined as $k_{xz} = \frac{1}{2}(\langle u'^2 \rangle + \langle w'^2 \rangle)$, where $u'$ and $w'$ are the fluctuating streamwise and vertical velocity components respectively, is presented in figure 3. In the plane aligned with the tower, shown in figure 3(a), TKE peaks are observed in the regions of high shear behind the blade tips and the hub. A strong reduction in TKE is observed behind the tower, which can be explained by the break-up of large-scale turbulent structures due to the presence of the tower, interrupting the turbulence cycle [14]. In figure 3(b), the TKE distribution in the off-tower plane shows strong peaks at the regions of high shear in the velocity field (figure 2b). Again, the effect of the tower and hub are not visible in this plane. The three-dimensionality of the wake becomes even more evident in figure 3(c) where the TKE profiles taken at $x = 0.41D$ from each spanwise plane are compared. The signature of the hub wake (i.e., the peak in TKE behind the hub) has completely disappeared in the off-tower plane, consistent with the weakening of the hub velocity peak seen in figure 2(c).

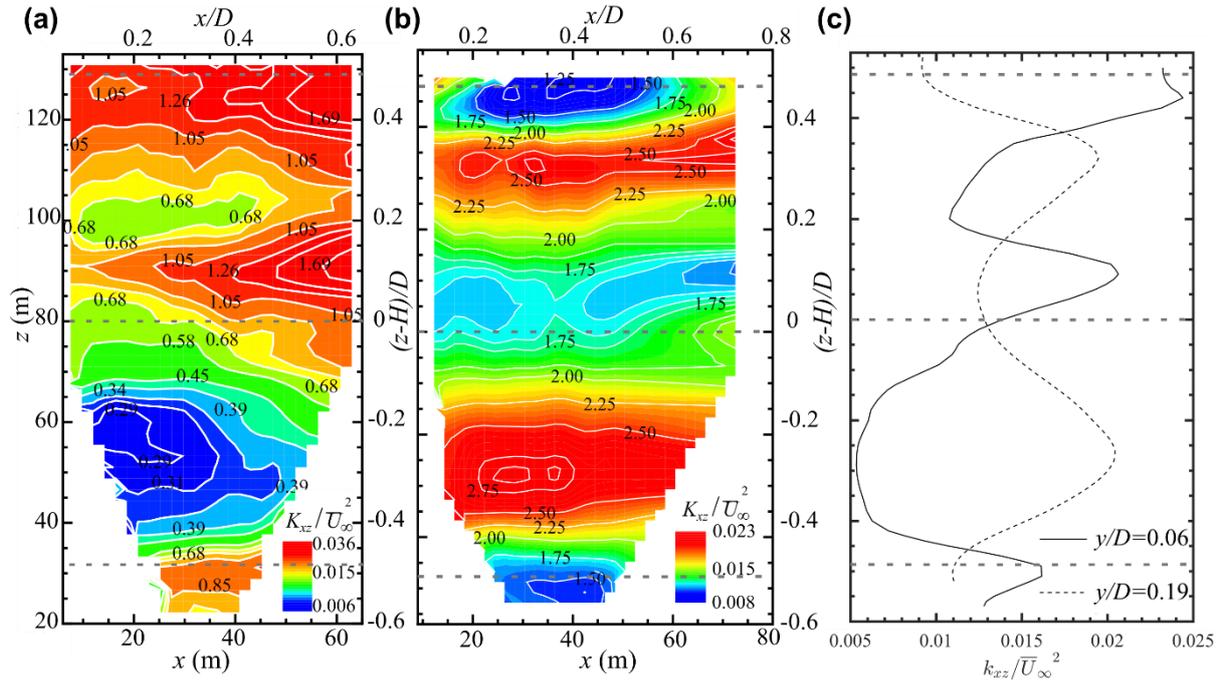

**Figure 3.** (a) In-plane turbulent kinetic energy field in the plane aligned with the tower. Dashed horizontal lines indicate hub and blade tip heights. (b) In-plane turbulent kinetic energy field in the plane offset from the tower. (c) Comparison of the in-plane TKE profiles taken at $x/D = 0.41$ for measurements at the spanwise locations $y/D = 0.06$ and $y/D = 0.19$.

*3.2. Instantaneous wake expansion*

Though the mean flow displays a clear velocity deficit characteristic of wake expansion, Dasari *et al* [13] revealed the existence of intermittent periods of wake contraction during utility-scale wind turbine operation. In the current study, instantaneous changes in wake expansion are analyzed from multiple views to develop a complete three-dimensional understanding of this behavior. From the view parallel to the flow, wake expansion is quantified by the wake ratio, $R_w = \bar{u}_{in}/\bar{u}_{out}$ where $\bar{u}_{in}$ is the spatial average of the velocity in the inner region indicated in figure 4(a) and $\bar{u}_{out}$ is the average over the outer region, as described in Dasari et al [13]. Note that the hub wake acceleration makes it impossible to quantify the wake expansion for the tower plane dataset using this method.

The instantaneous wake expansion can also be quantified using the view perpendicular to the flow (plan view) using the trajectories of the vortices shed from the blade tips. In the SLPIV images, vortices appear as dark regions, or voids, where the snowflakes have been expelled by the strongly rotating fluid. Image enhancement techniques are applied to extract these voids, and the processed images are stacked to reconstruct a quasi-three-dimensional visualization of the wake. The details of this procedure can be found in Abraham & Hong [15]. From the sample image in figure 1(f), it is clear that the bottom part of the wake is highly chaotic due to the interaction between the bottom blade tip vortices and the tower, so the wake expansion analysis focuses on the top tip where the characteristic tip vortex helix is visible. An envelope is fit to the reconstruction of the top tip vortex and compared to the geometric wake shape assuming no expansion extracted from a CAD model of the experimental setup (figure 4b). The vertical shift between the observed and geometric wake shapes ($\delta_{w,z}$) is used to approximate the wake expansion angle, defined as $\varphi_{w,z}$ [15].

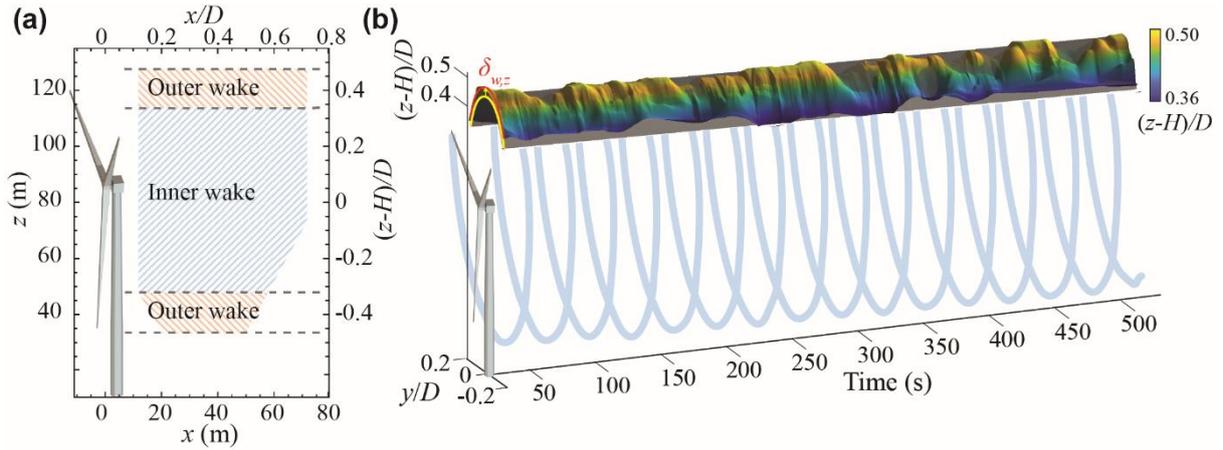

**Figure 4.** (a) Schematic showing the definition of the inner and outer wake used to calculate wake ratio, $R_w$, which defines wake expansion from the side view following Dasari et al [13]. (b) Schematic showing the wake envelope reconstructed from the plan view data. Difference between the experimental wake envelope (in colour) and the geometric envelope extracted from a CAD model (in grey) is used to define wake expansion from the plan view.

*3.2.1. Regional dependence.* Both Dasari et al [13] and Abraham & Hong [15] show that the wake expansion behaviour depends strongly on the turbine region of operation. In region 2, where the wind speed is relatively low and the blade pitch remains constant to maximize power generation, wake expansion changes less than in higher regions. For example, in the plan view dataset the standard deviation of $\varphi_{w,z}$ in regions less than or equal to 2 is 2.1°, compared to 3.6° in regions greater than 2 where the blade pitch is changing. In regions greater than 2, Abraham & Hong [15] show that wake expansion is strongly correlated with tip speed ratio (defined as $2U_\infty/\Omega D$, where $U_\infty$ is the incoming wind speed and $\Omega$ is the rotational speed or the rotor in rotations per minute) and blade pitch, with correlation values of 0.58 and -0.74 respectively. These parameters determine the thrust coefficient of the turbine, which in turn influences the axial induction, the parameter responsible for the reduced velocity and expansion of the wake. The effects of tip speed ratio and blade pitch on the flow experienced by the blades can be combined using the effective angle of attack, $\alpha_E = tan(2U_\infty/\Omega D) -$

$\beta - \beta_0$, where $\beta$ is the blade pitch and $\beta_0 = 2.1°$ is the blade pre-twist angle. Correspondingly, there is a strong relationship between wake expansion and $\alpha_E$, as shown in figure 5. Figure 5(a) shows a scatter plot of wake expansion angle from the plan view versus the effective angle of attack. These parameters have a correlation value of 0.65. In figure 5(b), the data is divided into periods of expansion ($\varphi_{w,z} > 0$) and contraction ($\varphi_{w,z} < 0$) and the histograms of $\alpha_E$ during these periods are compared. There is a clear separation between wake expansion and contraction at a value of $\alpha_E \approx 5.5°$. This plot can be compared to the histogram relating $R_w$ to $\alpha_E$ from the side view data in Dasari et al [13]. The plots are remarkably similar, though the trend is stronger in the plan view dataset (the correlation coefficient is 0.37 in the side view dataset). This may be due to the difference in amount of data from each region in the two datasets. The turbine spends 41% of the time in region 2.5 in the plan view dataset compared to only 34% in the side view dataset. This suggests the relationship with $\alpha_E$ may be stronger when the turbine operates in region 2.5. Additionally, the cut-off value of $\alpha_E$ between expansion and contraction is closer to 4° in the side view dataset. This discrepancy may be related to differences in the axial induction factor in the two datasets, which was neglected in the calculation of $\alpha_E$ due to the large uncertainty involved in its estimation.

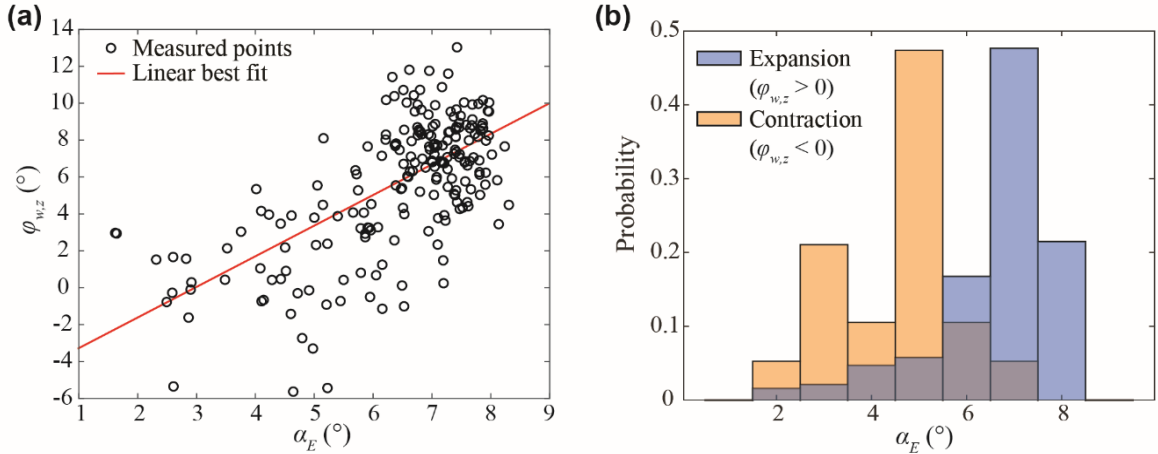

**Figure 5.** (a) Scatter plot showing the relationship between the effective angle of attack ($\alpha_E$) and the instantaneous wake expansion angle ($\varphi_{w,z}$), including a linear least squares best fit line to highlight the trend. (b) Histogram comparing $\alpha_E$ during periods of wake expansion versus periods of wake contraction.

*3.2.2. Region 3 operation.* When the data is further conditionally sampled for periods where the turbine is operating in region 3, more interesting phenomena are revealed. In region 3, the turbine relies on pitch control to regulate the loading on the structure, so the blade pitch changes continuously. A comparison between wake expansion from the plan view and this changing blade pitch, $d\beta/dt$, reveals a clear relationship (correlation value of 0.38), as shown in figure 6. Notably, periods of strong wake contraction are observed when $d\beta/dt < 0$. The physical explanation this phenomenon was proposed by Dasari et al [13], who observed an extremely strong correlation between blade pitch and the strain on the tower and blades. They propose that, when the blade pitch changes, the tower and blades are deflected backwards into the wake, inducing a vortex ring state that causes wake contraction. This physical understanding is supported by figure 6(b) which shows that contraction is more likely to occur when the blade pitch is decreasing. This plot is also consistent with the results from the side view reported in Dasari et al [13]. In this case, the relationship with $d\beta/dt$ is stronger in the side view dataset, likely because the turbine operates in region 3 59% of the time in the side view dataset relative to 29% in the plan view. As demonstrated in figure 7, the relationship between strain and pitch is strongest when the turbine is operating in region 3.

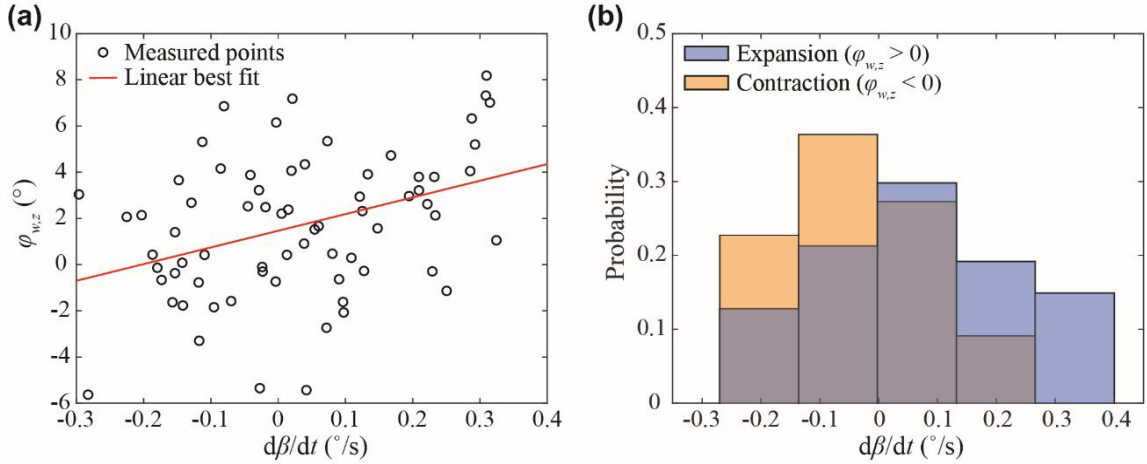

**Figure 6.** (a) Scatter plot showing the relationship between the blade pitch gradient ($d\beta/dt$) and the instantaneous wake expansion angle ($\varphi_{w,z}$) when the turbine is operating in region 3, including a linear least squares best fit line to highlight the trend. (b) Histogram comparing $d\beta/dt$ during periods of wake expansion versus periods of wake contraction in region 3.

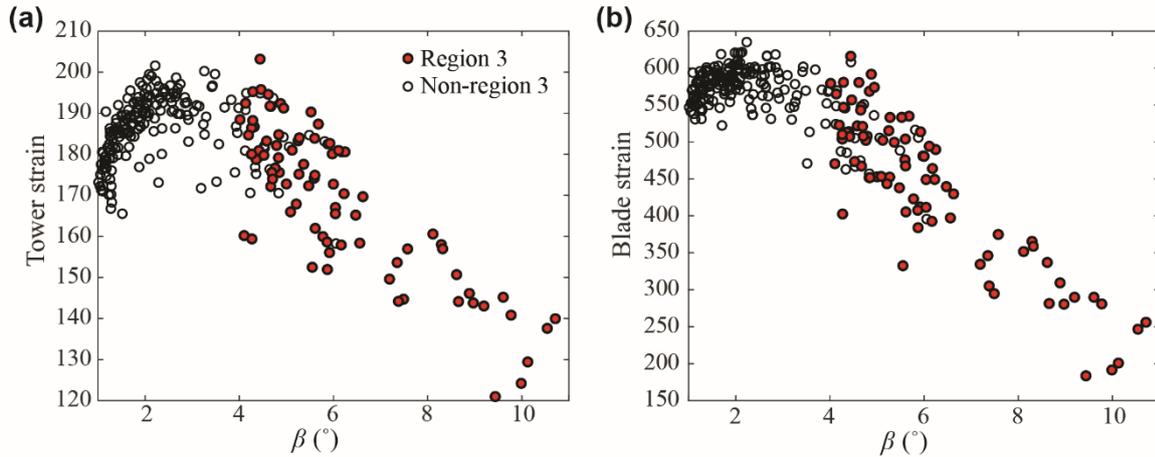

**Figure 7.** (a) Relationship between blade pitch and tower strain measured at the base of the tower when the turbine is operating in region 3 versus other regions. (b) Relationship between blade pitch and blade strain measured at 44.4% span of the blade (21.6 m from the blade root) on the high pressure side when the turbine is operating in region 3 versus other regions.

## 4. Discussion and conclusion

Through super-large-scale flow visualization and SLPIV, we explore the near wake of a utility-scale wind turbine from multiple views. The mean flow reveals the effect of the hub and tower on the wake. A region of high-speed flow is visible behind the hub where the axial induction is reduced, and increased blockage from the tower causes a region of low-speed flow below the hub. These features are not visible in the off-tower plane, demonstrating the limited spanwise influence of the hub and tower. The in-plane TKE further demonstrates peaks in turbulence in the high-shear regions behind the hub and blade tips. A reduction in TKE is observed behind the tower where large-scale turbulent structures are broken up. These features also disappear in the off-tower plane, leaving only the peaks in TKE at the regions of high shear near the blade tips.

Instantaneous changes in wake expansion, first observed by Dasari et al [13], are also investigated in more detail from two views, one parallel to the flow and one perpendicular. This phenomenon is strongly dependent on the region of operation of the turbine. In regions less than or equal to 2, wake expansion is relatively constant. In regions greater than 2, the degree of wake expansion depends on the effective angle of attack of the blades, related to the thrust coefficient. In region 3, periods of wake contraction are observed when the blade pitch is decreasing. This phenomenon can be explained by the strong relationship between blade pitch and structural strain on the turbine. When the blade pitch changes, the turbine deflects into its own wake, inducing wake contraction.

The utility-scale data presented here can be used to improve models of the near-wake by highlighting the effect of the hub and tower by revealing the significance of instantaneous changes in turbine operation. The unprecedented resolution also makes the data suitable for simulation validation. Additionally, the instantaneous behaviours described here occur in response to changes in turbine operational parameters that are readily available to the turbine controller. Therefore, the understanding of these behaviours can be incorporated into advanced control algorithms and future wind farm designs.

**Acknowledgements**

This work was supported by the National Science Foundation CAREER award (NSF-CBET-1454259), Xcel Energy through the Renewable Development Fund (grant RD4-13) as well as IonE of University of Minnesota. We also thank the students and the engineers from St. Anthony Falls Laboratory, including S. Riley, B. Li, Y. Wu, Y. Liu, J. Tucker, C. Ellis, J. Marr, C. Milliren and D. Christopher for their assistance in the experiments